\documentclass[preprint,showpacs,preprintnumbers,amsmath,amssymb]{revtex4}

\usepackage{graphicx} 
\usepackage{dcolumn} 
\usepackage{bm} 

\begin{document}

\preprint{}

\title{Multi-wall carbon nanotubes as quantum dots}

\author{M.\ R.\ Buitelaar}
\author{A.\ Bachtold}
\altaffiliation[Present address ]{TUDelft, 2628 CJ Delft, Netherlands.}
\author{T.\ Nussbaumer}
\author{M.\ Iqbal}
\author{C. Sch{\"o}nenberger}
\email{Christian.Schoenenberger@unibas.ch}
\homepage{www.unibas.ch/phys-meso}
\affiliation{Institut f\"ur Physik, Universit\"at Basel,
  Klingelbergstr.~82, CH-4056 Basel, Switzerland }


\begin{abstract}
We have measured the differential conductance $dI/dV$ of individual
multi-wall carbon nanotubes (MWNT) of different lengths.
A cross-over from wire-like (long tubes) to
dot-like (short tubes) behavior is observed.
$dI/dV$ is dominated by random conductance fluctuations
(UCF) in long MWNT devices
\mbox{($L=2\dots 7$\,$\mu$m)}, while
Coulomb blockade and energy level quantization are
observed in short ones \mbox{($L=300$\,nm)}.
The electron levels of short MWNT dots
are nearly four-fold degenerate (including spin) and
their evolution in magnetic field (Zeeman splitting) agrees
with a $g$-factor of $2$. In zero magnetic field
the sequential filling of states
evolves with spin $S$ according to
$S=0\rightarrow 1/2\rightarrow 0 \dots$.
In addition, a Kondo enhancement of the conductance is
observed when the number of electrons on the tube is odd.
\end{abstract}

\maketitle


Carbon nanotubes (NTs) are excellent model systems to study the
electronic properties of low-dimensional conductors
\cite{Schoenenberger1}. Transport and scanning probe measurements
on metallic single-wall nanotubes (SWNT, diameter \mbox{$d \sim
1$\,nm}) have revealed that these are one-dimensional conductors
($1$D) with long mean free paths,
\mbox{$l_{mfp}>1$\,$\mu$m}\cite{Tans,Bockrath,Bachtold1}. In
contrast, multi-wall carbon nanotubes (MWNT, outer diameter
\mbox{$d \sim 15$\,nm)} \cite{SpringerReview}, which are composed
of a set of coaxial NTs, were found to be disordered. The
transport regime varied from $2$D-diffusive in some MWNTs
\mbox{($l_{mfp}\lesssim 10$\,nm)} \cite{Langer,Bachtold2}, to
quasi-ballistic in others \mbox{($l_{mfp}\sim 100$\,nm)}
\cite{Schoenenberger2}. Most SWNT devices, even very long ones
with lengths \mbox{$L\sim 1$\,$\mu$\,m}, have displayed
single-electron tunneling effects \cite{Kouwenhoven1} with
conventional Coulomb blockade oscillations and a quantization of
the electron states (particle-in-a-box) at \mbox{$T \lesssim
10$\,K} \cite{Tans,Bockrath}. This demonstrates that the SWNTs
were only weakly coupled to the leads in these experiments. The
very fact that transport occurs through discrete electron states
implies that the corresponding molecular orbitals are
phase-coherent and extend over long distances, a remarkable result
for a $1$D electron system. Very recently, highly transparent
contacts to SWNTs could be realized \cite{Liang}. The physics of
these systems proved to be very rich and ranges from devices
dominated by higher-order co-tunneling processes (like e.g. the
Kondo effect) at intermediate contact transparencies $T\sim
0.1-0.5$ \cite{Nygard} to open ballistic systems with
transparencies approaching unity \cite{Liang}.

While there are many examples of SWNT quantum dots,
little effort has been gone into the investigation
of MWNTs as such systems.
Given the larger size,
the corresponding smaller energy scales
and, most notably, the intrinsic scattering,
it is not obvious that these experiments will
lead to the same results. As we will show below it turns out that
in many respects they do. Since the level spacing $\delta E$ scales
inversely with length, quantum dot features will be most pronounced in short
devices. We have therefore studied MWNTs with short
\mbox{$300$\,nm} inter-electrode spacing
We report
here measurements of one such device which showed pronounced
quantum dot features. For comparison we have also investigated
the linear-response conductance
of a very long MWNT  \mbox{($7.4$\,$\mu$m)}.

\begin{figure}
\includegraphics[width=120mm]{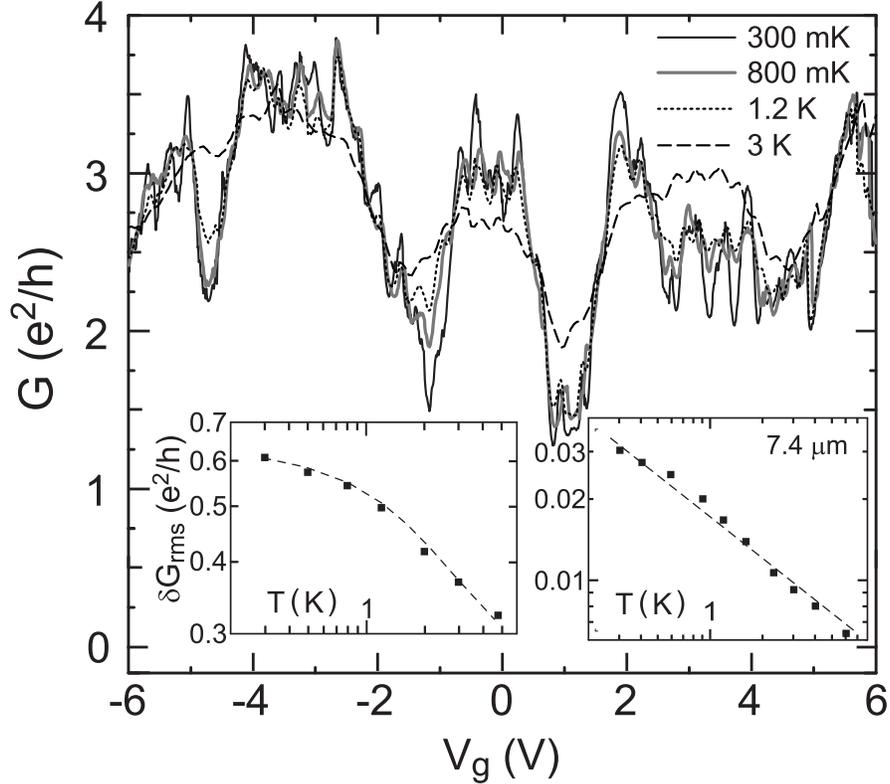}
\caption{\label{Fig1}
Linear response conductance $G$ as a function of
gate voltage $V_g$ (backgate) at temperatures between
\mbox{$280$\,mK} and \mbox{$3$\,K} for a MWNT
with contacts separated by \mbox{$300$\,nm}.
Lower left inset: the
corresponding root-mean square $\delta G_{rms}$ of the
conductance fluctuations $\delta G=G-\langle G\rangle$.
Lower right inset: as a reference,
$\delta G_{rms}$ of a very long MWNT device
with inter-electrode spacing of \mbox{$7.4$\,$\mu$m}.
The temperature dependence obeys
$\delta G_{rms} \propto T^{-1/2}$ (dashed line).
}
\end{figure}
Figure~\ref{Fig1} shows the dependence of the conduction $G$ of a \mbox{$300$\,nm}
long MWNT device as a function of gate voltage (backgate).
Details on device fabrication
can be found in Ref. \cite{Schoenenberger2}.
Large and reproducible fluctuations of order $e^2/h$ develop
in $G$. Note, that the average conductance is quite large,
i.e. \mbox{$\langle G\rangle > 2e^2/h$}, and nearly temperature
independent.
The root-mean square $\delta G_{rms}$ of the conductance
fluctuations $\delta G=G-\langle G\rangle$
is displayed in the lower left inset as a function of
temperature $T$.
For disordered wires random
variations of $G$ as a function of magnetic field or Fermi energy
(which is changed here by the gate) are usually assigned to
universal conductance fluctuations (UCF).
These fluctuations depend on the specific scattering potential but
are universal in the sense that their amplitude at {\em zero}
temperature is of order $e^2/h$, regardless of the sample size and
degree of disorder. At finite temperature, self-averaging reduces
$\delta G_{rms}$. For a wire which is $1$D with respect to the
phase-coherence length $l_{\phi}$ it is given by: $\delta G_{rms}
= \sqrt{12}(e^2/h)(l_{\phi}/L)^{3/2}$, where $L\gg l_{\phi}$ is
the wire length \cite{Lee}. When $l_{\phi}$ becomes of the order
of $L$, there is a cross-over to the universal value, $\delta
G_{rms} = 0.73e^2/h$ \cite{Lee}. The cross-over appears in the
measurement at \mbox{$T\alt 1$\,K} and $\delta G_{rms}$ $\approx
0.6e^2/h$ at \mbox{$T_0=280$\,mK}, in good agreement with theory.
At this temperature the average conductance amounts to $2.8e^2/h$.
The saturation of $\delta G_{rms}$ close to the universal limit
suggests conduction through one phase coherent unit. This in turns
implies that \mbox{$l_{\phi} \agt L$}.

As a comparison, we also show $\delta G_{rms}(T)$ of a long MWNT
with \mbox{$L=7.4$\,$\mu$m} in the lower right inset of Fig.~\ref{Fig1}
For this long nanotube device, $l_{\phi}/L\ll 1$, which
is reflected in the much smaller
amplitude of $\delta G_{rms}$.
$\delta G_{rms}=0.03e^2/h$ at $T_0$
corresponds to \mbox{$l_{\phi}=320$\,nm}.
This is consistent with the estimate of $l_{\phi}$
for the short MWNT
and also agrees with $l_{\phi}$ deduced from
magneto-conductance experiments of other MWNT
samples \cite{Schoenenberger2}.
The dephasing mechanism can be determined from the measured
temperature dependence of $l_{\phi}$ . Since $\delta G_{rms}$
follows a power-law in $T$ over more than one decade with an
exponent close to $-1/2$ (line in the lower right inset of
Fig.~\ref{Fig1}), $l_{\phi}(T) \propto T^{-1/3}$. This is the behavior
expected for a quasi-$1$D conductor. The main
contribution to phase-breaking is caused by electron-electron
collisions with small energy transfers, a process known as Nyquist
dephasing \cite{Altshuler}.

In terms of $l_{\phi}$ the long NT is a wire, whereas
the short one is a dot for which Coulomb blockade (CB)
may be important. We now estimate the
single-electron charging energy $U_C=e^2/C$ \cite{convenience}
for the latter and compare it with
the base measuring temperature \mbox{$T_0=0.28$\,K}.
The dominant contribution to the electrostatic capacitance $C$
of short MWNTs is due to the contacts, leading to
\mbox{$C\sim 400$\,aF} (see below). Hence, \mbox{$U_C\sim 4.6 $\,K},
i.e. $U_C/kT_0\sim 17$, and CB is therefore
expected.
This, however, does not seem to be the case:
firstly, the average conductance $\langle G \rangle$ is
nearly temperature independent (also over a larger temperature
range than shown in the graph); secondly, the conductance is large
$\langle G \rangle=2.8e^2/h$; and thirdly, $G$ lacks the
periodicity in gate voltage expected for a weakly coupled quantum
dot. We attribute the absence of clear CB features to a large
contact transparency. With increasing gate voltage from $0$ to
\mbox{$14$\,V}, however, the average conductance decreases from $2.8e^2/h$
to $\sim 2e^2/h$, possibly reflecting a decreasing contact
transparency. This assumption is supported by the following
measurement which displays clear
Coulomb blockade features
in the large gate-voltage regime.
\begin{figure}
\includegraphics[width=100mm]{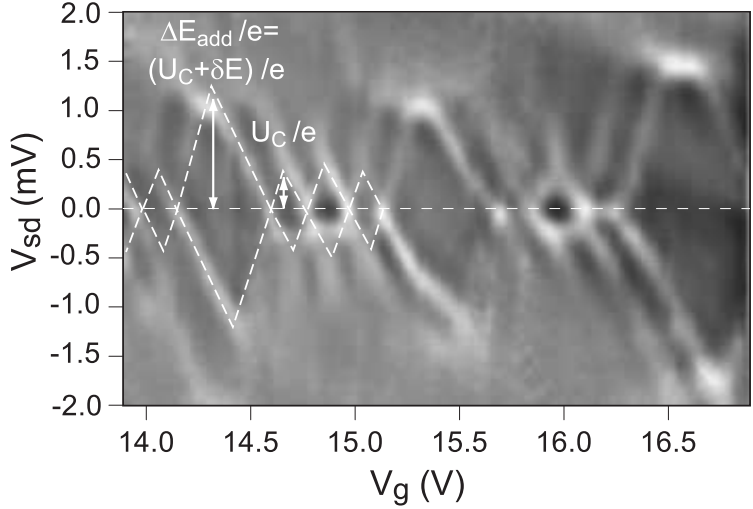}
\caption{\label{Fig2}
Greyscale representation of the
differential conductance as a function
of source-drain ($V_{sd}$) and gate voltage ($V_g$)
at \mbox{$280$\,mK} (lighter = more conductive).
The average 2-terminal conductance is high ($ \sim 2e^2/h$),
nevertheless clear traces of Coulomb blockade are observed.
The pattern of a large `diamond' followed by $3$ smaller ones
suggest a (nearly) $4$-fold degeneracy (including spin) of the
single-electron dot states. $\Delta E_{add}$, $U_C$, and $\delta E$
denote the addition energy, the charging energy
$U_C=e^2/C$, and the single-electron level spacing, respectively.}
\end{figure}
Fig.~\ref{Fig2} shows a grey-scale representation of $dI/dV$
versus gate voltage $V_g$ (horizontal) and transport voltage
$V_{sd}$ (vertical) in the regime \mbox{$V_g=14\dots 17$\,V}.
Although the overall two terminal conductance is still large, i.e.
$G\sim 2e^2/h$, the `remainders' of CB are clearly visible. The
most striking observation is a sequence of a large low-conduction
CB `diamond' followed by $3$ smaller ones (best seen on the left).
The diamonds are highlighted by dashed lines in the figure. The
size of the CB-diamonds reflects the magnitude of the addition
energy $\Delta E_{add}$, which measures the difference in chemical
potential of two adjacent charge states of the dot
\cite{Kouwenhoven1}. In the constant interaction model (C
independent of $N$) $\Delta E_{add}=U_C+\delta E$. If all the
single-electron levels would repel each other (only 2-fold spin
degeneracy) and $\delta E\sim U_C$, an alternating sequence of
small and large CB diamonds would be expected. Staring from an
even filling number, $\Delta E_{add}=U_C+\delta E$ for the first
added electron (large diamond) and to $U_C$ for the second one
(small diamond) \cite{Nygard}. The sequence of one large diamond,
followed by three smaller ones of approximate equal size, which is
observed here, suggests that the degeneracy of the states is not
$2$, but rather $4$ (including spin). From the size of the
diamonds we obtain  \mbox{$\delta E=0.8$\,meV} and
\mbox{$U_C=0.4$\,meV}, the latter corresponding to \mbox{$C =
400$\,aF}. The total capacitance $C$ is the sum of the gate
capacitance $C_g$ and the contact capacitances $C_l$ (left) and
$C_r$ (right). All three parameters can be deduced from the
diamonds. We obtain: \mbox{$C_g=1$\,aF} and
\mbox{$C_{l,r}=260,140$\,aF}.

The level spacing $\delta E$ of an ideal metallic SWNT
is given by  $\delta E=hv_F/2L$, where
$v_F$ is the Fermi velocity \cite{SpringerReview,Dresselhaus}.
This holds for an undoped NT.
Recently, it has been found that MWNTs are substantially
hole-doped by the environment \cite{Krueger}.
As a consequence more than the ideally
expected $2$ modes participate in transport. Hence,
$\delta E=hv_F/ML$, where $M>2$ is the number of
$1$D subbands. Taking \mbox{$v_F=8\cdot 10^{5}$\,m/s} \cite{Fermivelocity},
the measured value of \mbox{$\delta E=0.8$\,meV}
corresponds to $M\approx 14$, in good agreement with
Kr\"uger {\it et al.} \cite{Krueger}.
An estimate of the lifetime broadening $\Gamma$
can be obtained from the measured width of the Coulomb peaks and
yields \mbox{$\Gamma \approx 0.25$\,meV}.

The observed $4$-fold degeneracy can be explained by a specific
property of the graphite sheet (graphene). In the simplest
tight-binding band-structure calculation all $1$D-bands are
twofold degenerate (not including spin) \cite{Dresselhaus}.
This degeneracy can be traced back to
the presence of two C-atoms per unit cell, each contributing
with one valence orbital. This, so-called
\mbox{K-K$^{\prime}$}-degeneracy has not been observed before,
although it is supposed to be a {\em generic} feature of graphene.

\begin{figure}
\includegraphics[width=130mm]{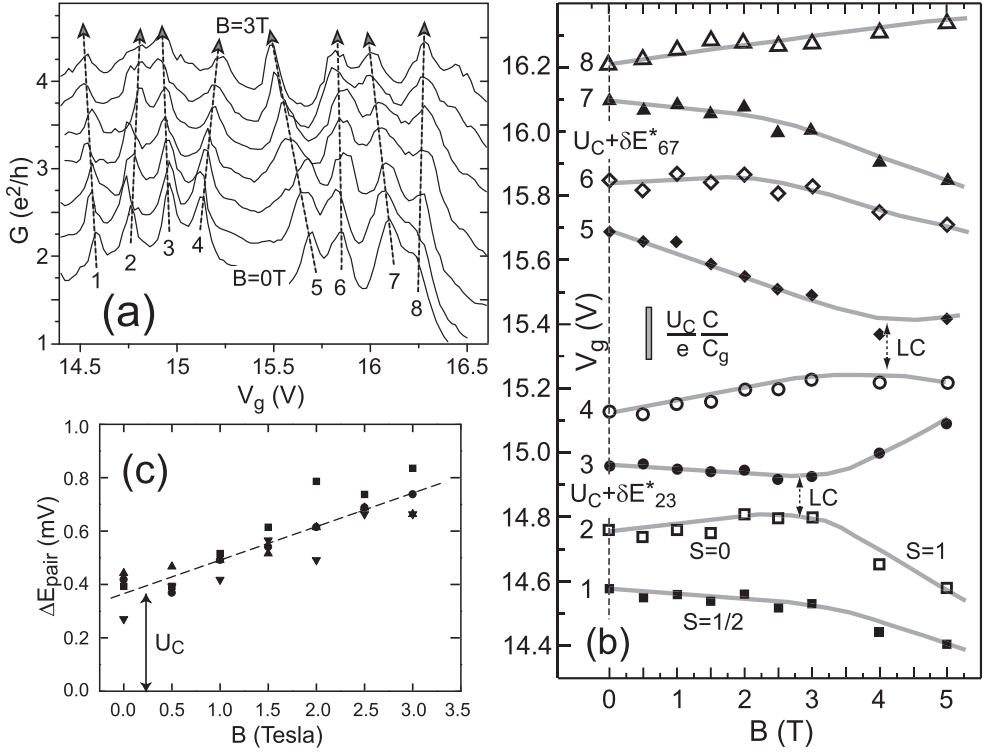}
\caption{\label{Fig3}
(a) Linear-response conductance $G$ as a function of
gate voltage $V_g$ for different magnetic fields
\mbox{$B=0\cdots 3$\,T} (vertically offset for clarity).
The evolution of the conductance peaks are highlighted by dashed lines.
(b) Peak positions versus \mbox{$B=0\cdots 5$\,T}. Curves are
guides to the eye and LC denotes level-crossings.
(c) Magnetic field dependence of the
addition energy $\delta E_{pair}$ deduced from the
separation of adjacent peaks involving electrons
on the same orbital (the pairs $1\leftrightarrow 2$
$3\leftrightarrow 4$, $5\leftrightarrow 6$, and
$7\leftrightarrow 8$). The dashed line (least-square fit)
corresponds to the Zeeman energy with
Land$\acute{e}$ factor $g=2.1$.}
\end{figure}
To explore this scenario further we have also studied the
gate-voltage shifts of the linear-response conductance peaks as a
function of a perpendicular magnetic field $B$, Fig.~\ref{Fig3}.
The difference between the positions of adjacent peaks can
also be related to the addition energy: $\Delta V_g=C\Delta
E_{add}/eC_g$. Figure~\ref{Fig3}a shows the evolution in small magnetic
field \mbox{$B\leq 3$\,T}. Adjacent peaks are seen to shift in
opposite directions. This is the behavior of a ground-state
whose spin alternates as $S=0\rightarrow 1/2\rightarrow 0\cdots$.
This, however, contradicts the assumed $4$-fold
degeneracy for the following reason: In the presence of a magnetic
field the energy of an electron depends on its spin due to the
Zeeman effect which lowers the degeneracy from $4$ to $2$ (only
Zeeman energy assumed). For $N=2$, the two electrons are thus expected to
occupy different orbitals with {\em parallel} spins. Actually, this
would already be expected for $B=0$, because of
exchange-correlation (Hund's rule would favour the spin-triplet
with total angular momentum $S=1$) \cite{Tarucha}.
The spin should therefore evolve as $S=1/2\rightarrow
1$ for $N=1\rightarrow 2$. Experimentally, however, the first $2$
electrons have opposite spins and are thus added to the
same obital state. This discrepancy can only be resolved, if the
assumed $4$-fold degeneracy is not `exact', i.e. there are pairs
of states which lie close together with spacing $\delta
E^{\star}$. The pairs themselves are spaced by
$\delta E > \delta E^{\star}$.
A detailed study of the peak evolutions (Fig.~\ref{Fig3}b) reveals
that this is indeed the case.
$\Delta E_{add}$ at $B=0$ of the
$2\rightarrow 3$ ($6\rightarrow 7$) transition
is clearly larger than the
$1\rightarrow 2$ and $3\rightarrow 4$ ($5\rightarrow 6$
and $7\rightarrow 8$) ones. We obtain \mbox{$\delta
E^{\star}_{23}\approx 0.1$\,meV} and \mbox{$\delta
E^{\star}_{67}\approx 0.18$\,meV}, on average \mbox{$\delta
E^{\star}\approx 0.14$\,meV}. We have also verified that the energy
shifts agree with the Zeeman term for electrons occupying the same
orbital. We plot in Fig.~\ref{Fig3}c
the corresponding addition energies as a function of $B$.
A best fit of the data to $U_C + g{\mu}_B B/e$ , where
$\mu_B$ is the Bohr magneton and  $g$ the Land$\acute{e}$
factor, is shown as a dashed line
and yields $g=2.1\,\pm\,0.3$. This value is consistent with
$g= 2.0$ for graphite and with previous measurements of $g$ for a
SWNT \cite{Tans,Cobden}.

In high magnetic field levels cross. Two crossings (LCs)
are seen in Fig.~\ref{Fig3}b. At \mbox{$\approx 3$\,T},
for example,
the spin-up of the first orbital crosses the spin-down of the
second, giving rise to an $S=0\rightarrow 1$ transition.
A similar crossing is not seen in the upper part. On the
one hand, this is due to the larger $\delta E^{\star}$.
On the other hand, there is also a magnetic-field dependence
of the orbitals which increases $\delta E^{\star}$ at higher fields.

We find a pattern that repeats every $4$th electron
due to an apparent {\em pairing} of orbital states. We believe that
this pairing is related to the $K-K^{\prime}$-degeneracy.
The splitting $\delta E^{\star}$ is proposed to be caused
by hybridization via the contacts, which are strongly coupled
to the NT. We expect a level `repulsion', in size comparable to
the life-time broadening, which we estimated to be
\mbox{$\Gamma\approx 0.25$\,meV}.
This is in fair agreement
with \mbox{$\delta E^{\star}\approx 0.14$\,meV}. In high magnetic
field the intrinsic $K-K^{\prime}$-degeneracy should be lifted
which enhances the level separation further. This may explain why the
$S=0\rightarrow 1$ transition is not observed for the upper quartet
in Fig.~\ref{Fig3}b. Finally, the fact that $S=0$ for $N=2$ at
$B=0$ is only consistent with Hund's rule if the exchange
energy $E_{X} < \delta E^{\star}$, yielding an upper
bound for $E_X$ of \mbox{$0.14$\,meV}.

\begin{figure}
\includegraphics[width=90mm]{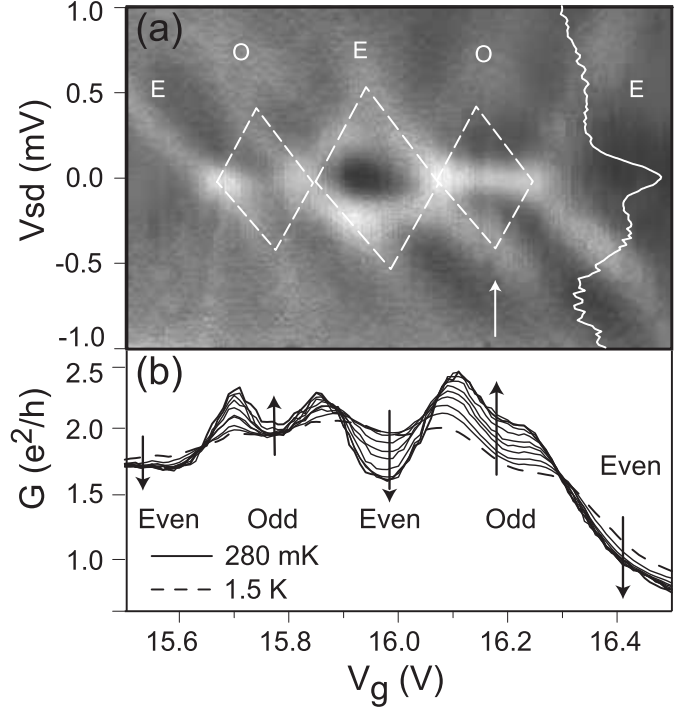}
\caption{\label{Fig4} (a) Greyscale representation of $dI/dV$ as a
function of source-drain ($V_{sd}$) and gate voltage ($V_g$) at
\mbox{$280$\,mK} (lighter = more conductive). Within regions
marked as $O$ ($E$) the number of electrons on the tube is odd
(even). The horizontal features (best seen in outer right diamond)
are caused by the Kondo effect. (b) Temperature dependence of the
linear-response conductance. The arrows indicate directions of
decreasing temperature. } \label{fig4}
\end{figure}
Another interesting manifestation of the electron spin on the
electronic transport can be seen in the gate region between $15.5$
and \mbox{$16.5$\,V}. We have measured this part with improved
accuracy and it is shown in Fig.~\ref{Fig4}. Fig.~\ref{Fig4}b
shows the linear-response conductance. In the valleys marked as
$E$ (even filling) the conduction decreases with decreasing
temperature while it increases in the valleys marked as $O$ (odd
filling). Contrary to what one might expect from normal CB, a high
conduction `ridge' around \mbox{$V_{sd}=0$\,V} develops in the
latter (best seen in the right most diamond). These observations
can be understood with the Kondo model
\cite{Kouwenhoven2,Nygard,Goldhaber}. When the number of electrons
on the tube is odd and the coupling to the leads is sufficiently
strong a spin singlet can form between the spin polarized tube and
electrons in the leads. This results in a resonance in the
density-of-states at the Fermi energy (i.e. the Kondo resonance).
The width of the Kondo resonance reflects the binding energy of
the singlet which is usually described by a Kondo temperature
$T_K$. The conductance is expected to increase logarithmically
with decreasing temperature in the centers of the ridges below
$T_K$. Following $G$ as a function of temperature at \mbox{$V_g =
16.2$\,V} we indeed find a logarithmic dependence between
\mbox{$280$\,mK} and \mbox{$1$\,K}. At temperatures well below
$T_K$ the conductance is expected to saturate at a maximum value
of $2e^2/h$. This is called the unitary limit. In our case,
however, no saturation has been observed down to \mbox{$280$\,mK}.

The Kondo effect is expected to be suppressed by a small bias
voltage across the tube of the order of $\pm k_B T_K /e$. The
ridge at $V_g = 16.2$ V has a width of $\sim$ 0.2 meV which would
correspond to $T_K$ = 1.2 K (see curve in Fig.~\ref{Fig4}a).
This is roughly in agreement with the
onset of the logarithmic increase of $G$ below $\sim$ 1 K.  An
additional prediction is the disappearance of the Kondo resonance
in a magnetic field. The high conductance ridge indeed broadens
and disappears above \mbox{$\sim 1.5$\,T}. Simultaneously, the
Coulomb blockade diamonds are recovered.

We acknowledge W. Belzig, G.\ Burkard, D. Cobden,
R. Egger and J. Nyg\aa rd for
discussions. We thank L. Forr\'o for the MWNT material and J.
Gobrecht for providing the oxidized Si substrates.
This work has been supported by the Swiss NFS.





\begin{thebibliography}{99}
\bibitem{convenience}
For convenience we use $U_C=e^2/C$ instead
of $E_C=e^2/2C$.
\bibitem{Fermivelocity}
The Fermi velocity of $8 \cdot 10^5$ m/s is an upper limit since
the average $v_F$ might be lower for a system doped into several
subbands.
\end{thebibliography}
\end{document}